# BEAM CURRENT MONITORS IN THE NLCTA*

Christopher Nantista and Chris Adolphsen

*Stanford Linear Accelerator Center, P.O. Box 4349, Stanford, California 94309, USA*

*Abstract*

The current profile along the 126 ns, multi-bunch beam pulse in the Next Linear Collider Test Accelerator (NLCTA) is monitored with fast toroids (rise time ~1 ns). Inserted at several positions along the beam line, they allow us to track current transmission as a function of position along the bunch train. Various measurements, such as rise time, current, width, and slope, are made on the digitized signals, which can be corrected in software by means of stored frequency response files. The design and implementation of these devices is described.

## I. INTRODUCTION

The Next Linear Collider Test Accelerator [1] at SLAC runs with an electron beam bunched at the X-band frequency of 11.424 GHz. A single pulse of ~126 ns duration contains a train of over 1400 bunches. This bunch train must be brought intact through the injector [2], a phase space tailoring chicane, and the accelerator proper, into a spectrometer and beam dump. A beam loading compensation scheme [3] is used to achieve uniform acceleration in the structures, minimizing energy variation along the pulse as it travels through the steering and focusing lattice.

As part of the beam instrumentation system of the NLCTA, six current monitoring toroids are incorporated at various locations along the beam line. These supplement the current information available from an insertable faraday cup and a series of beam position monitors.

While the latter are more numerous, the current transmission data acquired from them consists of a scalar value reflecting only the charge density at the leading edge of the long beam pulse. The toroid pickups allow us to view the temporal structure of the beam current (above the bunching scale). By comparing signals at different locations, for example before and after the chicane, we can determine not only where we might be losing current but also from which part of the bunch train it is being lost. This is crucial as we test our beam loading compensation scheme.

## II. TRANSFORMER

The heart of our beam current monitoring system is the toroidal Fast Current Transformer manufactured by Bergoz (01170 Crozet, France). Developed in collaboration with Klaus Unser at CERN, this device was designed specifically for viewing charged particle bunches in accelerators. The "rad-hard" model we use has an inner diameter of 1.1", slightly larger than our beam pipe. The azimuthal magnetic induction of the beam current links a twenty-turn coil around the toroid core, giving a 1:20 transformer ratio. The core of cobalt/molybdenum amorphous alloy ribbon interleaved with nickel/iron crystalline alloy was designed to optimize frequency response (~1ns rise time) and suppress ringing. Fifty ohms of resistance is embedded within the toroid and an SMA connector on the outer wall allows for signal pick-up. The basic circuit is illustrated in Figure 1 below. When connected to an oscilloscope with 50Ω impedance, the transformer yields a nominal signal amplitude of

$$V_{\text{sig}} = \frac{I_{\text{beam}}}{20 \text{ turns}} \times \left( \frac{1}{50\Omega} + \frac{1}{50\Omega} \right)^{-1} = 1.25\Omega \times I_{\text{beam}}.$$

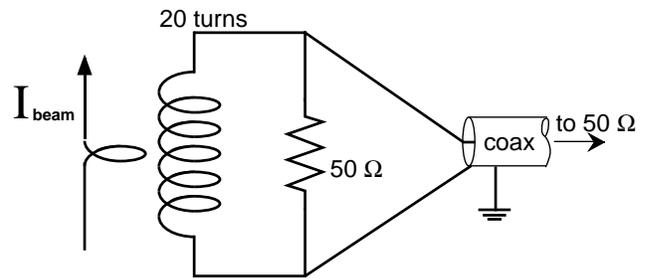

Figure 1. Circuit diagram for fast toroid current monitor.

## III. HOUSING

In order for the transformer to "see" a net current passing through its center, the wall current imaging the charged particle beam must be diverted around the outside of the device. The beam pipe section and aluminum housing on which and in which the toroid is installed to achieve this end are shown in Figure 2 in a partly disassembled view.

The electrical continuity of the beam pipe is thus interrupted by an inch-long insulating ceramic gap. A thin coating of kovar deposited on the inner surface of the ceramic provides a finite conductivity to bleed off any build-up of charge deposited by the beam. The edges of the ceramic are metalized and brazed to the beam pipe to maintain vacuum integrity, and a small bellows is included to either side of the gap to relieve stress.

Around this structure, a mechanically engineered aluminum shell is assembled which supports the toroid and provides a conducting path for the wall current. The housing clamps to the steel beam pipe just beyond each bellows. An inner cylinder is brought into contact around the toroid, and an outer cylinder encloses some cabling, supports two cable connectors, and provides rigidity. Spring-loaded pins suspend the toroid around the ceramic. The transformer signal is brought through the inner cylinder to an SMA-to-type N adapter in the outer shell. To the other type N connector is soldered a flat wire looping beween the toroid and the vacuum chamber. This can be pulsed to verify the operability of the current monitor.

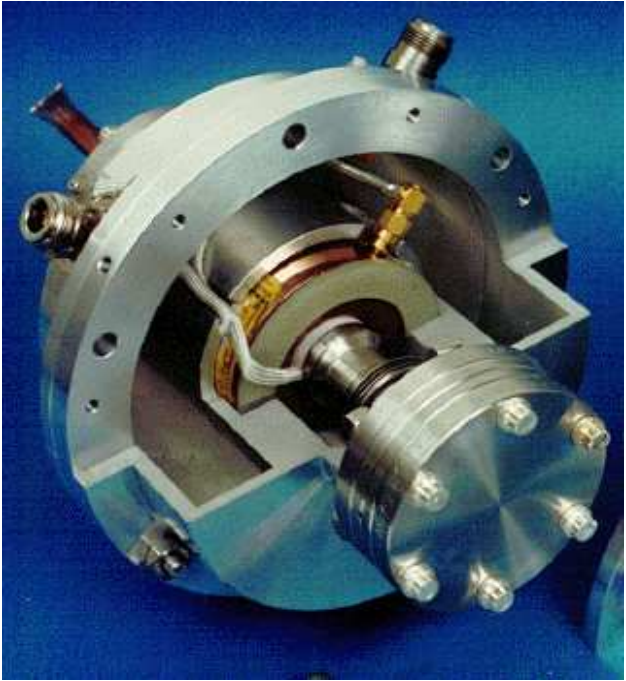

Figure 2. Toroid and housing installed on a beam pipe section with one quadrant of the housing removed.

## IV. READOUT

The fast toroid signals are carried by low-loss heliax cables out from the accelerator enclosure to a rack near the klystron gallery. There they are fed through a high bandwidth multiplexer into a 250 MHz (1 gigasample/second) Hewlett Packard 1428a Digitizing Oscilloscope in a VXI crate. These instruments are controlled with an embedded, HPUX, multi-user processor, which is networked to a terminal in the NLCTA control room. The scope is run using an external trigger synched with the beam. Software routines written with HP VEE (Visual Engineering Environment) graphical programming language are used for sending commands, receiving digitized data, signal processing and display.

## V. CALIBRATION

In an initial test of our toroid assembly in the SLC linac, the signal displayed significant ringing. This seems to have been due to capacitance and inductance in the circuit formed by our housing. To further study its response and try to ameliorate this effect, we fitted the current monitor with an aluminum rod center conductor supported between specially designed coaxial cable adapters, as shown in Figure 3.

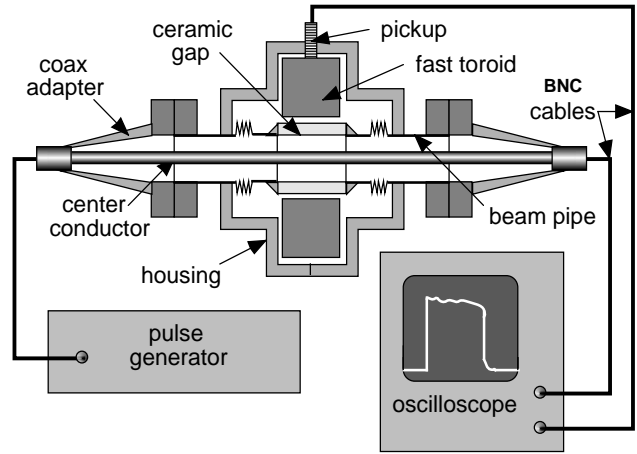

Figure 3. Beam current monitor test and calibration setup.

An HP 8510C Network Analyzer connected between one port and the pickup, with a load on the other port, showed the response function of the current transformer to be fairly flat out to about 1.3 GHz, where it drops off 15 dB. This is consistent with the manufacturer's rise time specification.

A variable width pulse generator and an oscilloscope were also used to study the device. Filling the cavity formed around the bellows with microwave absorber reduced the ringing from sharp pulses, but it was not considered suitable for permanent inclusion in a high radiation environment. What was found to improve the worst of our current monitors was the sanding down of spacers and application of silver paint to the edges of the inner cylinder halves to assure good electrical contact between them.

To remove, as far as possible, any remaining imperfection in their response, each of our beam current monitors was calibrated in the following way.

Electrical pulses were sent along the center conductor to simulate an electron beam. The transmitted pulses (essentially identical to the input pulses) and toroid pickup signals were recorded simultaneously with our two-channel digitizing scope. The Fourier transform of the latter waveform could then be divided by that of the former to give us the effect of the device on the true signal.

Because the Fourier transform of a square pulse has nodes at intervals of the inverse width, a response function generated from a single such pulse yields spurious noise (division by nearly zero) at these frequencies. Attempts to correct a pulse of a different width with such a function causes gross distortions. It was thus necessary to use many pulses of various widths for each calibration, generally about fifteen ranging from 1 ns to greater than a microsecond. The digitized waveforms were eight microseconds long, yielding a spacing of 0.125 MHz in our frequency spectra. The FFT's of the through signals and those of the toroid signals were added before dividing the two sums.

The resulting response functions were then stored in files. For any signal $h(t)$ recorded by the beam current monitor, a corrected signal $f(t)$ could thereafter be obtained with the stored file $G(\omega)$ from the relationship

$$f(t) = IFFT\left\{\frac{FFT\{h(t)\}}{G(\omega)}\right\}.$$

For greater speed, it is sometimes preferable to run our current monitors without implementing this correction. A scalar factor, also determined with our coaxial setup, is then used to convert output voltage to axial current. A typical effect of using the fast Fourier transform correction is to decrease a ~7 ns rise time to ~6 ns. Signal droop is not significant for our pulse length. We measured the L/R time constant of the toroid to be as large as 35 microseconds.

It should be noted that the performance of these current transformers deteriorates drastically in high magnetic fields (>~500 guass). In our injector, where the focusing solenoid field reaches two kilogauss, we had to substitute a homemade, 100 turn toroid on a G-10 core.

## VI. MEASUREMENTS

As previously mentioned, the HP VEE programming software is used with our beam current monitoring system. Through its direct I/O capability, we select the desired current monitor or monitors from the multiplexer, set up the digitizing scope, and acquire waveforms. The signal processing we've encoded not only allows for the above FFT correction to be applied, but also performs a number of useful measurements on the pulses. These include beam current amplitude, pulse width, rise time, fall time, slope of the pulse top, rms variation across the pulse top from a linear fit, and integrated charge in the pulse. These measurements can be sent to the CORRELATIONS PLOT program in our control system and correlated with other variables.

The waveforms and measurement results are displayed on a convenient user interface panel. We can run in one-shot or "live" mode. In the latter mode, the screen is updated automatically every few seconds (including time for data processing). A feedback loop adjusts the oscilloscope range according to the measured signal amplitude, and the two scope channels allow us to monitor two beam current signals simultaneously on the same display. Figure 4 shows our display panel with toroids before and after the chicane indicating about 77% beam transmission.

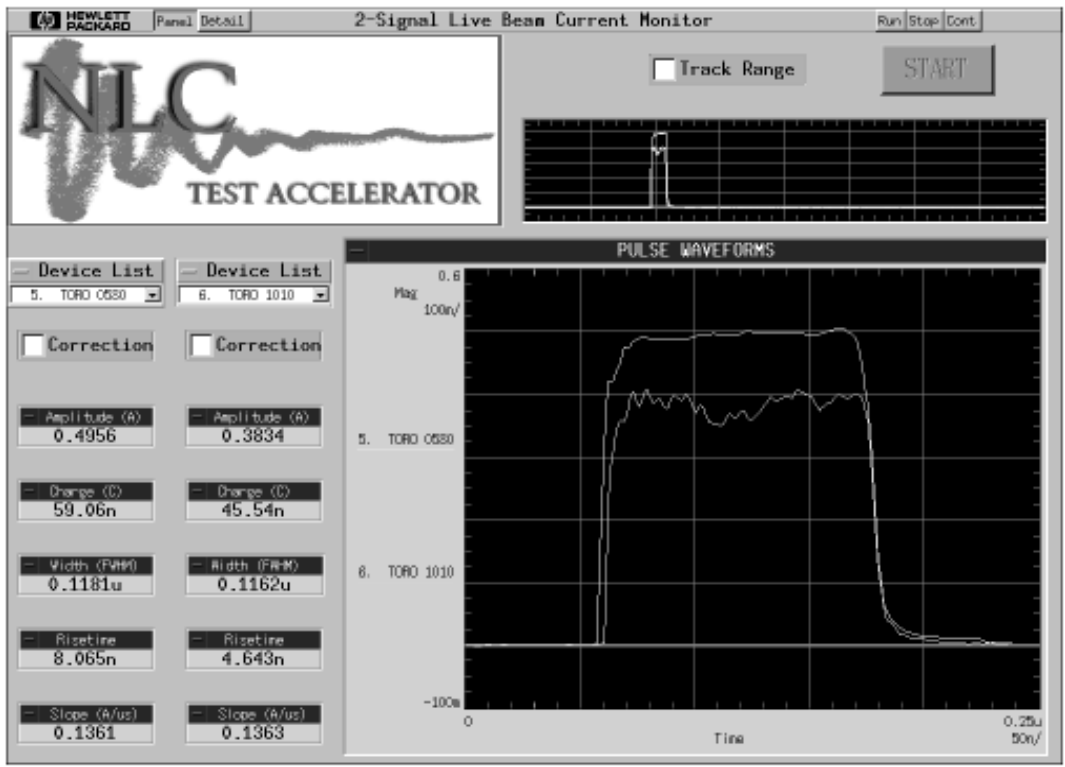

Figure 4. Current monitor two-signal live display panel.